\documentclass[galaxies,article,moreauthors,pdftex,10pt,a4paper]{mdpi}

\firstpage{1} 
\articlenumber{x}
\doinum{10.3390/------}
\pubvolume{xx}
\pubyear{2018}
\copyrightyear{2018}
\history{Received: date; Accepted: date; Published: date}

\pdfoutput=1

\usepackage{amsmath}  
\usepackage{amssymb,marvosym}  
\usepackage{color}

\preto{\abstractkeywords}{\nolinenumbers}

\newcommand{\aap}{{A\&A}}

\newcommand{\aapr}{{A\&A Rev.}}

\newcommand{\apj}{{ApJ}}
\newcommand{\aj}{{AJ}}
\newcommand{\apjl}{{ApJ}}
\newcommand{\apjs}{{ApJS}}

\newcommand{\araa}{{ARAA}}
\newcommand{\bain}{{Bulletin of Astronomy in the Netherlands}}

\newcommand{\mnras}{{MNRAS}}

\newcommand{\nat}{{Nat}}

\newcommand{\pasj}{{PASJ}}

\newcommand{\ssr}{{SSRv}}

\newcommand{\procspie}{{SPIE}}

\newcommand{\rxj}{{RX\,J1713etc}}
\newcommand{{\ism}}{interstellar medium}
\newcommand{{\snr}}{supernova remnant}
\newcommand{{\snrs}}{supernova remnants}
\newcommand{{\pwn}}{pulsar wind nebula}
\newcommand{{\pwne}}{pulsar wind nebulae}
\newcommand{{\dsa}}{DSA}
\newcommand{\arcmin}{{$^{\prime}$}}
\newcommand{\arcsec}{{$^{\prime\prime}$}}
\renewcommand{\deg}{{$^\circ$}}

\Title{Practical aspects of X-ray imaging polarimetry of supernova remnants and other extended sources}

\Author{Jacco Vink $^{1,2}$\orcidA{}, Ping Zhou $^{3,4}$}

\AuthorNames{Jacco Vink, Ping Zhou}

\address{%
$^{1}$ \quad GRAPPA \& Anton Pannekoek Institute for Astronomy, University of Amsterdam, Science Park 904, 1098 XH Amsterdam, The Netherlands; j.vink@uva.nl\\
$^{2}$ \quad SRON, Netherlands Institute for Space Research, Sorbonnelaan 2, 3584 CA Utrecht, The Netherlands\\
$^{3}$ \quad Anton Pannekoek Institute for Astronomy, University of Amsterdam, Science Park 904, 1098 XH Amsterdam, The Netherlands; p.zhou@uva.nl\\
$^{4}$ \quad School of Astronomy and Space Science, Nanjing University, Nanjing 210023, China}
\corres{Correspondence: j.vink@uva.nl; Tel.: +31-20-525 7068}

\nolinenumbers

\abstract{
The new generation of X-ray polarisation detectors,  gas pixel detectors, which will be employed by
the future space missions IXPE and eXTP, allows for spatially resolved X-ray polarisation studies. 
This will be of particular
interest for X-ray synchrotron emission from extended sources like young
supernova remnants and pulsar wind nebulae. Here we report on employing a 
polarisation statistic that can be used to makes maps in the Stokes $I$, $Q$,
and $U$
parameters, a method that we expand by correcting for the energy-dependent
instrumental modulation factor, using optimal weighting of the
signal.
In order to explore the types of Stokes maps that can be obtained,
we present a Monte Carlo simulation program called {\em xpolim}, with which
different polarisation weighting schemes are explored.
We illustrate its use
with simulations of polarisation maps of young supernova remnants,
after having described the
general science
case for polarisation studies of supernova remnants, and its
connection to magnetic-field turbulence. We use  xpolim simulations to
show that in general deep, $\sim $2~Ms observations are needed to
recover polarisation signals, in particular for Cas A, for which in
the polarisation fraction may be as low as 5\%.}


\begin{document}


\nolinenumbers

\section{Introduction}
\label{sec:intro}

The polarisation of {electromagnetic} radiation is an important diagnostic tool to determine the nature of the radiation mechanisms as well as obtaining information on the
geometry and radiative transfer properties of radiation in and  near the source of radiation. 
 X-ray polarisation is in that sense no different, but has the added value that X-ray emission comes from high-energy sources with extreme conditions.
 For  young supernova remnants (SNRs) and pulsar wind nebulae (PWNe), X-ray polarisation holds the promise to determine
 the geometry and turbulent properties of the  magnetic fields, which determines the polarisation fraction and orientation of the magnetic fields, in case
 that the radiation is caused by X-ray synchrotron emission.
 These are also properties  that can be assessed through radio polarisation structures, but X-ray synchrotron comes from the highest energy electrons ($>$10~TeV, e.g., \cite{vink12}),
 which are generally short lived ($\lesssim$100~yr) and, therefore, sample plasmas  that are still near the sites where the electrons have been accelerated.
 Since acceleration of charged particles depends on the strength and the turbulence of the magnetic field, X-ray polarisation holds  a great promise in determining the
 conditions under which the electrons have been accelerated.

 X-ray polarisation studies have been done in the past, but without the ability of imaging polarisation structures.
 They typically made use of the linear polarisation dependence of Thomson/Compton scattering,
 or were based on scattering by a Bragg crystal.
  For extended sources the absence of imaging capabilities is far from ideal
 as  inside the sources there are likely to be regions with different polarisation properties,  diminishing or completely wiping out the overall polarisation fraction 
 that can be measured. Nevertheless, for the Crab nebula a net X-ray polarisation signal was measured with sounding rocket experiments~\citep{novick72} and
Bragg crystal polarimeters on board the OSO-8 satellite~\citep{weisskopf76,weisskopf78}. For the Crab nebula the overall X-ray polarisation fraction is $\Pi =19$\%
with a position angle of 154\deg~\citep{weisskopf78},
in agreement with optical and radio observations~\citep{oort56,mayer68}. However, more detailed maps in the radio~\citep{bietenholz90} and optical~\citep{moran16} show regions with much higher
polarisation fraction,
 and even evolution of polarisation as a function of time for the inner regions~\citep{moran16}. It is this level of imaging detail that has been lacking in X-ray polarisation observations so far,
 but this is going to change in the near~future. 

The ability to image X-ray polarisation has been brought about with the development of  
 polarisation sensitive gas pixel detectors (GPDs)~\citep{costa01,bellazzini07}, which makes
use of the photoelectric effect in a thin layer of gas, and the fact that the photoelectron ejection angle $\alpha$ is modulated as $\cos^2$ with respect to the polarisation vector of the X-ray emission.
The photoelectron creates a cascade of secondary free electrons, whose positions are recorded. From the location and shape of the electron cloud the original ejection angle has to be
reconstructed.

{
This new type of X-ray polarisation detectors 
 has led to new proposals for satellite based X-ray polarisation missions:
the {\em X-ray Imaging Polarimetry Explorer} (XIPE)~\citep{soffitta13} for ESA,  the {\em Imaging X-ray Polarisation Explorer} (IXPE)~\citep{weisskopf16} for NASA, and
the {\em enhanced X-ray Timing and Polarimetry Mission}
(eXTP)~\citep{extp16} for the Chinese space science program.
XIPE was considered for the ESA M4 science program, but it was recently announced that XIPE did not make the final selection\footnote{See \url{http://sci.esa.int/cosmic-vision/59796-esa-s-next-science-mission-to-focus-on-nature-of-exoplanets}.}. A decision that was to be expected, as NASA did select  the very similar IXPE  mission
for its {\em small mission} program. It~uses for example the same GPD detectors. IXPE  is expected to be launched in 2021/22.
Also eXTP uses this type of detector, but the eXTP mission combines X-ray polarisation instruments together with X-ray timing detectors. eXTP faces this year (2018) a final
review, and if successful could be launched around 2025\footnote{See \url{http://www.sciencemag.org/news/2018/03/china-unveils-plans-x-ray-satellite-probe-most-violent-corners-universe}.}.
}

The purpose of this article is to explore the imaging capabilities of X-ray imaging polarimeters based on GPDs, with a special focus on young SNRs for which the  expected
polarisation signals are weaker than for PWNe.
We start with explaining the use of the photon-by-photon statistics method (e.g., \cite{kislat15}),  which has obvious advantages over
the more traditional fitting of modulation curves (e.g., Fig. 10 in \cite{muleri10}) when it come to  polarisation  imaging, as no  extraction regions need to predefined, and
one can just make maps in Stokes Q and U, and its derivates (polarisation fraction $\Pi$, and polarisation angle).
We then describe a Monte Carlo code, which helps to define optimum observation strategies for young SNRs and PWNe, such as observation times and energy band optimisation.
It also allows us to assess what the effects are of the point spread function on the polarisation signals.
We then explain what  X-ray polarisation measurements can teach us about particle acceleration in young SNRs, and what kind of signal to expect.

Some earlier results were already reported in~\citep{deona17}, but here we provide new results and a new scheme for correcting for the instrumental modulation factor.


\section{Measuring X-ray polarisation with gas-pixel-detectors on a photon by photon statistics base}
\label{sec:statistics}

Gas-pixel-detectors (GPDs) to be used by the IXPE \citep{weisskopf16} and eXTP \citep{extp16} satellite experiments do not measure
directly the polarisation of the electric vector, but rather the direction of the
photon-electron, which is ejected from an atom in the gas-filled detector, if ionised by an X-ray photon.
The angle  $\alpha$ of the direction of the photon-electron is $\cos^2\alpha$ modulated with respect to the associated electric-field vector of the photon. 

In X-ray polarisation studies, traditionally the polarisation is measured by binning the measured angles, $\alpha_i$, and then
fitting the function 
\begin{equation}\label{eq:mu}
M(\alpha)=\left[1- \Pi + 2\Pi \cos^2(\alpha -\alpha_0)\right] \sin(2\alpha),
\end{equation}
to the binned data, using a $\chi^2$ minimisation method.

It is easy to apply (\ref{eq:mu}) for non-imaging devices, as one simply bins the measured events per angle $\alpha$ and fit the histogram.
However, with GPDs one would like to make optimum use of the imaging capabilities. Using (\ref{eq:mu})  would then require predefining
an image region of interest, and bin the $\alpha$ for all events in that region. Or, if the statistics is sufficient, do this for every pixel. But
doing the binning and minimisation for each pixel separately  is computationally expensive.

However, there is an easier method, which does not require binning the data for ejection angle $\alpha$.

The method is laid out in \citet{kislat15} and we explain it here, but using slightly different notations. 
As~estimators for the Stokes parameters $I$, $Q$, $U$ we use: 
\begin{align}\label{eq:qu_est}
\hat{I} \equiv &  \sum_{i=1}^{i=N} 1 = N\\\nonumber
\hat{Q}\equiv & 2 \sum_{i=1}^{i=N}  \cos (2\alpha_i), \\\nonumber
\hat{U}\equiv &  2 \sum_{i=1}^{i=N}  \sin(2\alpha_i).
\end{align}
The angle $\alpha_i$ will depend on the true polarisation direction $\alpha_0$ as $\cos^2(\alpha -\alpha_0)$. It is a matter
of taste whether one takes the average  (divide $\hat{Q}$ and $\hat{U}$ by $N$) or the plain sum as shown here.
In the absence of any polarisation signal the expectation values are $E\hat{Q}=E\hat{U}=0$. 

The equation is almost similar to the definition of the $Q$ and $U$ Stokes parameters, except that there is a factor 2, which 
accounts for the fact that not the electric-vector, but a $\cos^2$ modulated statistical realisation of it is measured.
This factor was omitted in the definition of $Q$ and $U$ in \citep{kislat15}, but there it accounted for when calculating the true
polarisation fraction. 

The {\em measured} polarisation fraction is  given by
\begin{equation}\label{eq:pi}
\Pi = \frac{\sqrt{ \hat{Q}^2 + \hat{U}^2}}{\hat{I}}.
\end{equation}
We use here the same symbol $\Pi$  as for the modulation amplitude (\ref{eq:mu}), because they are essentially the same. For
example the expectation value of $\hat{Q}$ for a signal that varies as $M(\alpha)$  (\ref{eq:mu}) is
\begin{align}
E(\hat{Q})=&\frac{2}{2\pi}\int_0^{2\pi}  \left[ (1-\Pi) + 2 \Pi \cos^2(\alpha -\alpha_0)\right] \cos(2\alpha)d\alpha 
=0 + 
\frac{1}{\pi} \int_0^{2\pi} \Pi
\left[1 + \cos(2\alpha -2\alpha_0) 
\right]\cos(2\alpha)d\alpha\\ \nonumber
=& 0  + \Pi \frac{1}{\pi} \int_0^{2\pi} \left[ 
\cos^2 (2\alpha)\cos(2\alpha_0) + \cos(2\alpha) \sin(2\alpha)\sin(2\alpha_0) 
\right]d\alpha\\\nonumber
= & \Pi \frac{1}{\pi} \left\{
\cos(2\alpha_0)\int_0^{2\pi}  \cos^2(2\alpha) d\alpha
+
\frac{1}{2} \sin(2\alpha_0)\int_0^{2\pi}  \sin(4\alpha) d\alpha\right\} = \Pi \cos(2\alpha_0).\nonumber
\end{align}
And similarly, $E(\hat{U})=  \Pi \sin(2\alpha_0)$.
Equation (\ref{eq:pi}) provides an estimate for the true polarisation factor, if there are no instrumental depolarisation effects.
How to correct for the instrumental effects will be discussed below.

The variance for $\hat{Q}$ and $\hat{U}$ can be directly estimated  assuming Poisson statistics:
\begin{align}\label{eq:qu_est}
\mathrm{Var}({\hat{Q}})\equiv &4 \sum_{i=1}^{i=N}  \cos^2 (2\alpha_i), \\\nonumber
\mathrm{Var}({\hat{U}})\equiv &4 \sum_{i=1}^{i=N}  \sin^2(2\alpha_i).
\end{align}
The expectation values  (see also \citep{kislat15}) in the absence of a polarisation signal are
 $E \mathrm{Var}({\hat{Q}})= E\mathrm{Var}({\hat{U}})=2N$.
Note that here we estimate the variances from the measured event data itself, whereas in
\citep{kislat15} one makes  use of  the expectation values for estimating the variances, assuming no polarisation is present.  

The advantage of using (\ref{eq:qu_est}) for X-ray {\em imaging} polarimetry are obvious: for each imaging pixel
$\hat{Q}$ and $\hat{U}$ and their variances can be easily calculated. If the signals are noisy
one can just rebin the $\hat{I}$, $\hat{Q}$ and $\hat{U}$ maps, because rebinning is summing pixels together, which just is equivalent
to calculating $\hat{Q}$ and $\hat{U}$ using more events.

\begin{figure}
\centerline{
  \includegraphics[trim=50 50 100 100,clip=true,width=0.6\columnwidth]{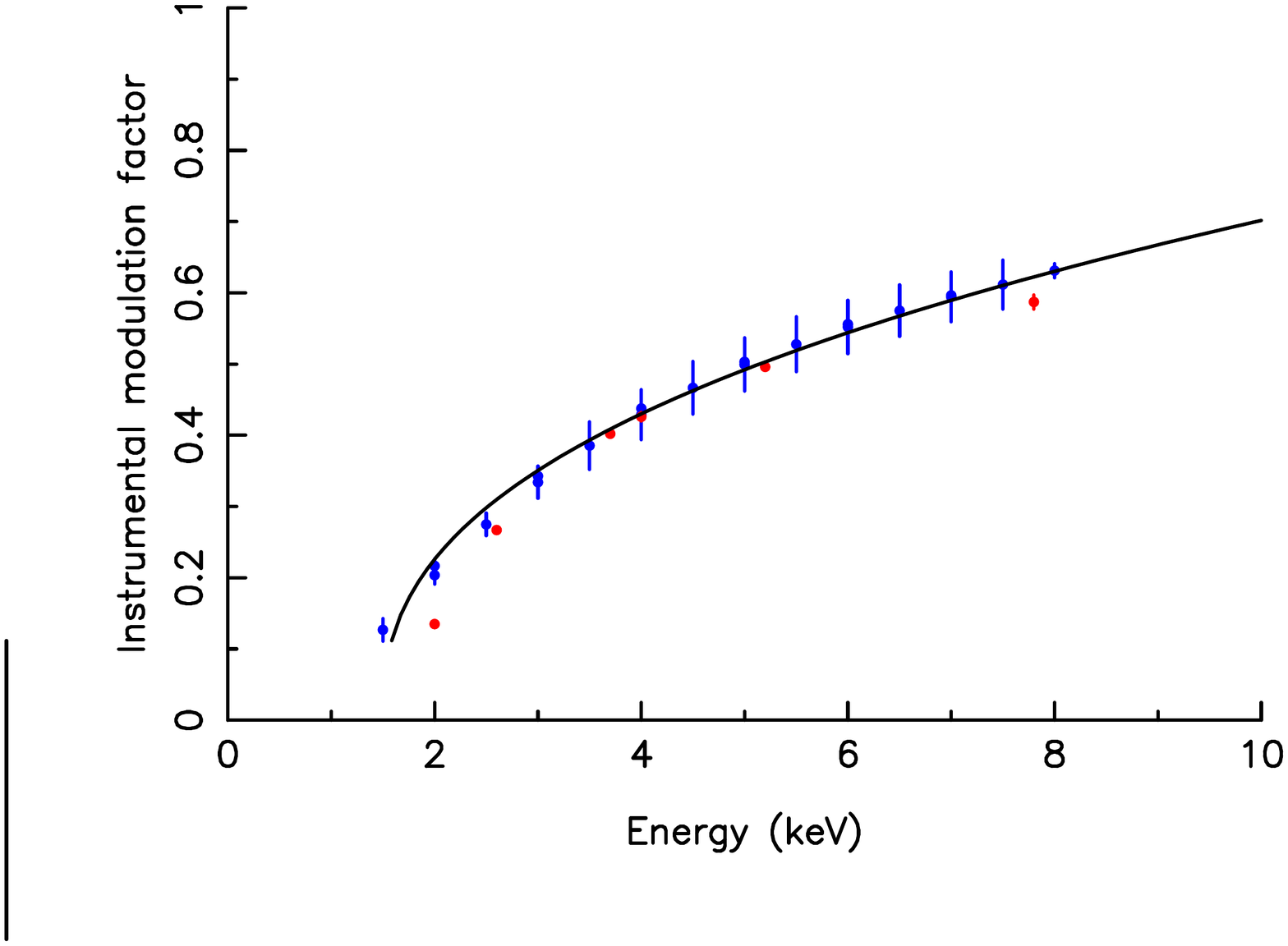}
  }
  \caption{\label{fig:modfactor}
    The modulation factor, $\mu$, as a function of energy, i.e. 
    the measured polarisation fractions for a 100\% polarised source,
    not correcting for instrumental
    effects.
    The red points are measurements taken from \citet{muleri10}.
    The blue points are based on the Monte Carlo code described in this text,
    and uses a Gaussian error $\Delta \alpha \propto \sqrt{E}$
    for the photo-electron ejection
    angle. The curved line is an analytical approximation for the
    modulation factor, used for the corrections on a event-by-event base.
  }
\end{figure}

However, the formalism does not take into account that the measured angles $\alpha_i$ contain errors.
These angles reduce the polarisation signal. If a hypothetical  source would be 100\% polarised $\Pi=1$
the measured polarisation signal will be $\Pi=\mu<1$, with $\mu$ the so-called modulation factor,
which is energy-dependent factor (Fig.~\ref{fig:modfactor}).

If $\hat{Q}$ and $\hat{U}$ would be determined in a narrow energy band an estimate of the  polarisation fraction of the source itself
is
\begin{equation}
\hat{\Pi}=\frac{1}{\mu} \frac{\sqrt{ \hat{Q}^2 + \hat{U}^2}}{\hat{I}}.
\end{equation}
In practice one may want to use all available events rather than selecting a narrow energy band, which suggests that
we correct each individual event with its energy-specific modulation factor $\mu_i$:
\begin{align}\label{eq:qu_est2}
\hat{Q}_\mathrm{corrected} \equiv & 2 \sum_{i=1}^{i=N} \frac{1}{\mu_i} \cos (2\alpha_i), \\
\hat{U}_\mathrm{corrected} \equiv &  2 \sum_{i=1}^{i=N}   \frac{1}{\mu_i}  \sin(2\alpha_i),\\ \nonumber
\mathrm{Var}(\hat{Q}_\mathrm{corrected} )\equiv & 4 \sum_{i=1}^{i=N} \left(\frac{1}{\mu_i}\right)^2 \cos^2 (2\alpha_i), \\
\mathrm{Var}(\hat{U}_\mathrm{corrected} )\equiv & 4\sum_{i=1}^{i=N} \left(\frac{1}{\mu_i}\right)^2 \sin^2(2\alpha_i). \nonumber
\end{align}
This would indeed provide an estimate of the source polarisation itself, but it has one disadvantage:
It is essentially a weighted sum, with $\frac{1}{\mu_i}>1$ as a weight. If the modulation factor is small,
i.e. the instrument/energy specific polarisation performance is poor, we give that event more weight.

Instead what we would like is to make a weighted sum that still provides a proper estimate of the true
polarisation factor, but that optimises the polarisation signal to noise ratio.
From a statistical point of view the optimum weighting scheme is to use inverse variance weighting which uses
weights $w_i=1/\sigma_i^2$ \citep{meier53,shahar17}.
Inspecting Eq.~\ref{eq:qu_est2} shows that if we were to use narrow energy bins the variance of the signal
scales with $1/\mu_i^2$, which implies that we take as weights
$w_i=\mu_i^2$.
With these weights applied the estimator that optimises the signal to noise  becomes
\begin{align}\label{eq:qu_est3}
\hat{I}_\mathrm{optimal}\equiv &   N \frac{\sum_{i=1}^{i=N} \mu_i}{\sum_{i=1}^{i=N} \mu_i^2}, \\
\hat{Q}_\mathrm{optimal}\equiv & 2  N \frac{\sum_{i=1}^{i=N} \mu_i \cos (2\alpha_i)}{\sum_{i=1}^{i=N} \mu_i^2}, \\
\hat{U}_\mathrm{optimal} \equiv &  2 N \frac{\sum_{i=1}^{i=N} \mu_i \sin (2\alpha_i)}{\sum_{i=1}^{i=N} \mu_i^2}, \\ \nonumber
\mathrm{Var}(\hat{Q}_\mathrm{optimal})\equiv & 4 N^2 \frac{\sum_{i=1}^{i=N} \mu_i^2 \cos^2 (2\alpha_i)}{\left(\sum_{i=1}^{i=N} \mu_i^2\right)^2}, \\\nonumber
\mathrm{Var}(\hat{U}_\mathrm{optimal})\equiv & 4 N^2 \frac{\sum_{i=1}^{i=N}\mu_i^2\sin^2(2\alpha_i)}{\left(\sum_{i=1}^{i=N} \mu_i^2\right)^2}. \nonumber
\end{align}
which if all $\mu_i$ would be identical would give exactly the same result as Eq.~\ref{eq:qu_est2}.
The multiplication factors $N$ (and $N^2$ for the variances) can be omitted, but serve here to make that the definition of the Stokes I  parameter,  $\hat{I}$,
remains identical to  Eq.~\ref{eq:qu_est}.
For convenience we drop the subscript "optimal", and  write $\Delta \hat{Q} \equiv  \sqrt{\mathrm{Var}(\hat{Q})}$ and $\Delta \hat{U} \equiv \sqrt{\mathrm{Var}(\hat{U})}$.

The estimators $\hat{Q}, \hat{U}, \mathrm{Var}{(\hat{Q})},\mathrm{Var}(\hat{U})$ contain all the information necessary. In practice one would like to quantify the polarisation signal in terms of the
polarisation fraction
\begin{equation}
\hat{\Pi} \equiv \frac{\sqrt{\hat{Q} ^2+ \hat{U}^2}}{\hat{I}},
\end{equation}
and position angle of the polarisation vector
\begin{equation}
\phi = \frac{1}{2} \arctan\left(\frac{\hat{U}}{\hat{Q}}\right).
\end{equation}
One has to be careful here when reporting these two quantities and their errors, as both quantities are based on a combination of $\hat{Q}, \hat{U}$ and are, therefore, correlated.
In the literature one often encounter the term polarisation intensity, which is simply the intensity in polarised light only $\hat{I}_\mathrm{pol}=\sqrt{\hat{Q} + \hat{U}}=\Pi \hat{I}$.

To find the signal-to-noise ratio of the polarisation measurement we calculate first the error in $\sqrt{\hat{Q}^2 + \hat{U}^2}$:
\begin{equation}
\Delta \left(\sqrt{\hat{Q}^2 + \hat{U}^2}\right) =
\left[  \left( \frac{\partial \sqrt{\hat{Q}^2 + \hat{U}^2} }{\partial \hat{Q}} \Delta \hat{Q} \right)^2 + \left( \frac{\partial \sqrt{\hat{Q}^2 + \hat{U}^2} }{\partial \hat{U}} \Delta \hat{U} \right)^2\right]^{1/2}=
\frac{\sqrt{ \hat{Q}^2(\Delta \hat{Q})^2  + \hat{U}^2(\Delta \hat{U})^2}}{\sqrt{\hat{Q}^2 + \hat{U}^2}}.
\end{equation}
The signal-to-noise ratio for the polarised intensity is therefore 
\begin{equation}\label{eq:significance}
S/N = \frac{ \sqrt{\hat{Q}^2 +\hat{U}^2}}{\Delta \left(\sqrt{\hat{Q}^2 + \hat{U}^2}\right)  }= \frac{\hat{Q}^2 + \hat{U}^2}{\sqrt{ \hat{Q}^2(\Delta \hat{Q})^2 + \hat{U}^2(\Delta\hat{U})^2}}.
\end{equation}
Note that the method advocated in \citep{kislat15} is to use for the variance the expectation value for the variances in absence of a polarisation signal.  Since in the absence of a signal
the expectation value for the variance is maximised the method in \citep{kislat15} is a little bit more conservative.

We consider estimates based on equations (\ref{eq:qu_est3}) as the preferred statistic for measuring polarisation. Note that one may use the weights/modulation factors based on
measured calibration results as depicted in Fig.~\ref{fig:modfactor}, but that in the future estimates for the weights $1/\mu_i^2$ may be based on
estimated errors for each individual event; for example on how easily $\alpha_i$ could be determined from the  electron cloud morphology.

Finally we point out that the method outlined here for making maps in $Q$ and $U$ can also be used for timing and spectral bins in order to make polarisation lightcurves and polarisation spectra.

\section{A Monte Carlo code for X-ray gas pixel detectors}
\label{sec:montecarlo}

The GPDs allow for the imaging of X-ray polarisation structures of the extended sources like PWNe and SNRs.  However, a problem we encounter for
SNRs is that the polarisation fraction may be low (see \S~\ref{sec:snrs}). Moreover, as we will discuss below, the X-ray synchrotron emitting regions are in some cases, like Cas A, confined to small
regions near the forward (and reverse) shocks, whereas from the rest of the SNR shell, non-polarised, thermal emission is emitted.
The ratio of X-ray synchrotron emission over non-thermal emission varies from SNR to SNR; for Cas A, Kepler's SNR, and Tycho's SNR  most of the X-ray emission is
thermal in nature, and the X-ray synchrotron emission is best studied in the 4-6 keV band, in which there is almost no line emission. For studies with GPDs that means
that the analysis has to be confined to this narrow band, which limits the statistics. 
For SN\,1006 (but also \rxj\ and Vela Jr)  the synchrotron emitting regions are broader, but also the X-ray synchrotron emission dominates the overall emission, certainly above
2~keV. Although these SNRs have a lower X-ray surface brightness, the larger energy bandwidth partially compensates for this.

\begin{figure}
  \includegraphics[width=\textwidth]{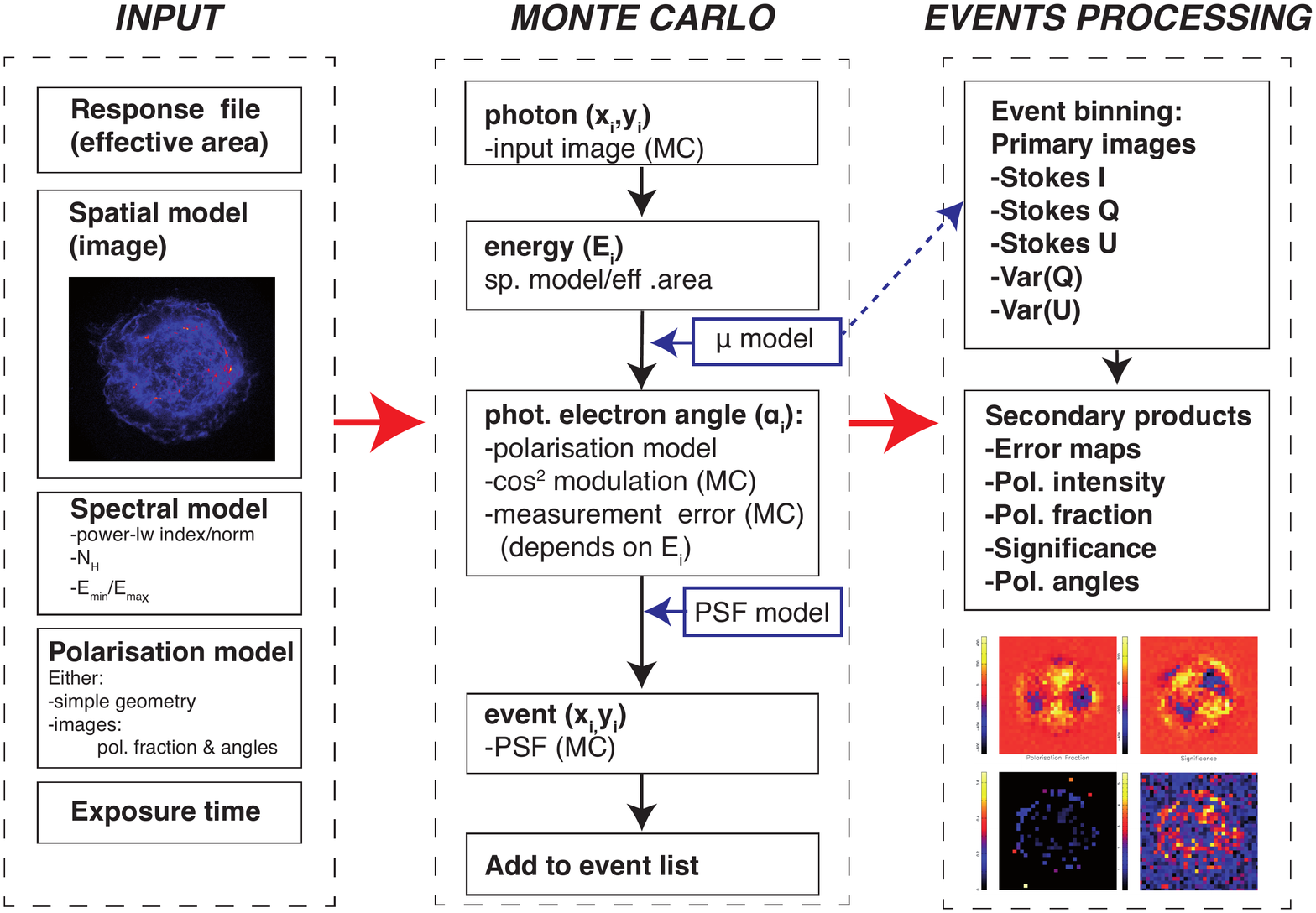}
  \caption{\label{fig:montecarlo}
  Schematic view of the {\em xpolim} code.}
\end{figure}

To study the effects of the energy band used, the low polarisation fractions to be expected based on the radio observations, and to study the effects of the point spread function (PSF)
on mixing photons from thermally emitting regions with non-thermally emitting regions, it is best to resort to Monte Carlo simulations of the expected signals. 
Moreover, a Monte Carlo code allows us to test the statistical methods explained in \S~\ref{sec:statistics}.
We set out to make a Monte-Carlo code to test imaging with the XIPE mission, but since IXPE and eXTP will carry similar detectors and telescopes, the results will be relevant  to all three missions.
The set-up of the Monte-Carlo code, named {\em xpolim}\footnote{
Currently {\em xpolim} is not publicly available, but an executable for MacOS  will be made available on request by the J. Vink. 
Depending on the demand and availability of funding for further development of {\em xpolim},  we will work toward a documented and portable version of the source code.
}
 is illustrated in Fig.~\ref{fig:montecarlo}.  

The code uses as an input a high resolution X-ray image of the object, for example a Chandra image of Cas A in the 4-6 keV band. Based on a spectral model (for now
the code only deals with power-law spectral models)  and the instruments effective area (mirror plus detector) the number of detect photons and their
energy distribution is predicted for a given observation time. The photon arrival directions are randomly drawn from the input image,
and based on a toy model of the polarisation direction and polarisation fraction a photon-electron angle $\alpha$ is assigned to the event.
The values $\alpha$ takes into account the $\cos^2 $ modulation based on the  physics of the photo-electric effect, and in addition a
random Gaussian error is added, which represents that the initial direction of the photo-electron cannot be accurately determined from the electron cloud.
Since the number of electrons scales with the energy as $N_\mathrm{pe} \propto E$, and the error on reconstruction is probably partially Poissonian in nature,
we assumed that $\Delta \alpha \propto \sqrt{N_\mathrm{pe}} \propto \sqrt{E}$. After some experimenting with the Monte-Carlo simulations we found that
\begin{equation}
\Delta \alpha = 5^\circ + 35^\circ \sqrt{\frac{E}{4~\mathrm{keV}}},
\end{equation}
gives a good fit to the measured modulation factor above 2~keV reported in \citep{muleri10} (see Fig.~\ref{fig:modfactor}).
We used this $\Delta \alpha$, but also its associated modulation factor, represented by the analytical function
\begin{equation}
\mu = 29.8 \left(E - 1.5~\mathrm{keV}\right)^{0.4} \%,
\end{equation}
for $E>1.5$~keV, otherwise $\mu=0$.
This functional form, which is used to estimate the actual polarisation signal using (\ref{eq:qu_est3}), is shown as a solid line in Fig.~\ref{fig:modfactor}.

After assigning a photoelectron angle to each event, the event itself is subject to a random realisation of the PSF, for which we used
the analytical description reported in \citep{fabiani14}, which is based on calibration measurements of the GPD in combination with X-ray optics.

The output of the {\em xpolim} program is an event list, which can then be processed according to the statistical method described in \S~\ref{sec:statistics}.
In fact, all three methods (no weighting, correction for $\mu$ and a weighted correction) are implemented, and can be tested.
The output itself consists of binned images of the Stokes $I$, $Q$, $U$ and their variances, as well as their derived products:
the polarisation fraction, its error, the polarisation detection significance (\ref{eq:significance}) and reconstructed polarisation angles.

Currently  spectral redistribution of the events based on a response file is not yet implemented, and the input spectral models are limited
to power-law distributions modified by a simple interstellar absorption model \citep{morrison83}.

 \section{The expected X-ray polarisation signals from shell-type supernova remnants}
 \label{sec:snrs}

\begin{figure}
\centering
 \includegraphics[trim=50 50 100 50,clip=true,width=0.6\textwidth]{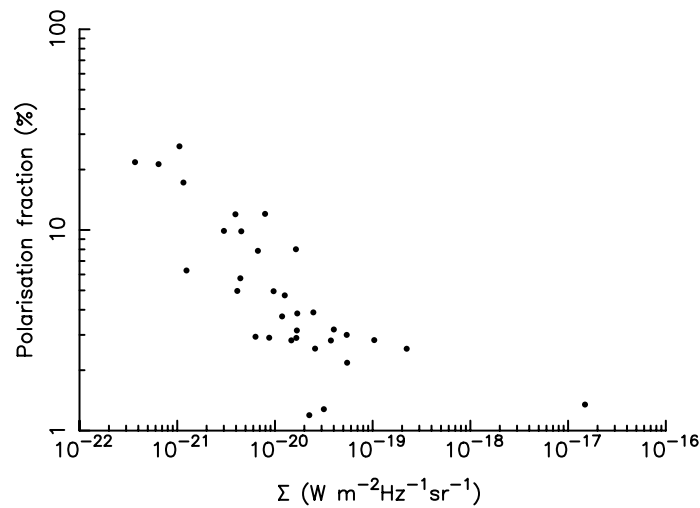}
  \caption{\label{fig:dickel}
    The radio polarisation fraction as a function of the radio surface
    brightness, taken from  \citep{dickel90} and based largely on the sample in
    \citep{milne75}. Surface brightness is here a
    proxy for size and/or age; low surface brightness remnants tend to be larger and older. 
    So this plot shows that young SNRs have low
    polarisation fractions (few \% to $\sim 20$\%).
    }
\end{figure}

\begin{table*}
  {\scriptsize
    \caption{\label{tab:radiopol}
      Radio polarisation measurements of young SNRs}
  \begin{tabular}{lrlllll}\hline\hline\noalign{\smallskip}
    Object & Typical $\Pi$ & Peak $\Pi$ & Orientation $B$ & $\lambda$&Remarks & Ref.\\
    \noalign{\smallskip}\hline\noalign{\smallskip}
    RCW 86 & 8\% &  15\% & radial & 22 cm& some regions $\Pi<3$\%&\cite{dickel01}\\
    SN\,1006 & 17\% & 60\% & mostly radial & 20 cm&peak not in X-ray rims &\cite{reynoso13}\\
    SN 1572 & 7\% & 25\% & radial/fine struct. (4\arcsec) & 6 cm& peak at limbs &\cite{dickel91}\\
    SN 1604  & 6\% & 12\% (?) & radial/fine struct. (20\arcsec)& 6 cm& &\cite{delaney02}\\
    Cas A & 5\% & $\sim$20\% & radial & 6 cm & outer plateau $\Pi\approx 9$\%& \cite{braun87b,anderson95b}\\
    G1.9+0.3& 6\% & $17\pm 3$\% & radial & 6 cm  & Faraday rotation?&\cite{dehorta14}\\
    \noalign{\smallskip}\hline
  \end{tabular}
  }
\end{table*}

The realisation that the radio emission from SNRs was caused by synchrotron emission from relativistic electrons, provided the first observational evidence for a link between cosmic-ray acceleration
and supernovae \citep{shklovsky54}. Since long the mechanism to accelerate the electrons  and protons to relativistic  energies is generally accepted to be diffusive shock acceleration (DSA) \citep[][for a review]{malkov01}.
Radio synchrotron emission is caused by accelerate electrons with a few GeV of energies, whereas the cosmic rays of Galactic
origin extend all the way up to the cosmic-ray "knee" at $3\times 10^{15}$~eV. To accelerate to these high energies, magnetic fields have to be considerably stronger
 than the average Galactic magnetic field strength of $\sim 5~\mu$G, and  the magnetic field needs to be highly turbulent.
In fact according to DSA theory the magnetic-field turbulence is created by the cosmic rays themselves, as they  diffuse ahead of the shock front. A process that has drawn
substantial theoretical attention, but for which the observational data is still scarce \citep[e.g.][]{bell04,marcowith16}
 
 The discovery of {\em X-ray} synchrotron emission from shell-type SNRs, first for SN\,1006 \citep{koyama95}, but later for virtually all SNRs younger than 1000-2000~yr \citep[][for a review]{vink12}
 showed that electrons can be accelerated to very high energies of $10^{13}-10^{14}$~eV, which requires in fact very high levels of turbulence and high shock velocities.
 The synchrotron cut-off photon energy for synchrotron radiation should occur around \citep{aharonian99}
 \begin{equation}
 h\nu_\mathrm{max}\approx 
 3 \left< \left(\frac{\delta B}{B}\right)^2\right>  
\left(\frac{V_\mathrm{sh}}{5000\ \mathrm{km\ s}^{-1}}\right)^2
 ~\mathrm{keV}.
 \end{equation}
This equation is derived by setting the acceleration rate equal to the energy-dependent synchrotron loss rate of the electrons. The electron synchrotron loss time for 10-100~TeV
is very short compared to the ages of SNRs:
\begin{equation}
\tau_\mathrm{syn}\approx 12.5 \left(\frac{B}{100\ \mu\mathrm{G}}\right)^{-2}\left(\frac{E_\mathrm{e}}{100\ \mathrm{TeV}}\right)^{-1}\mathrm{yr}.
\end{equation}
This short synchrotron loss time scale explains why X-ray synchrotron emission is only found close to the shock fronts that accelerate the particles. In fact, the
thickness of the X-ray synchrotron rims can be used to estimate the magnetic field strengths \citep{vink03a}, which appear to be $20-500~\mu$G \citep[][for an overview]{helder12b}.

The fact that young SNRs emit X-ray synchrotron implies that the magnetic fields are not very structured ($\delta B/B\sim 1$). Indeed, radio polarisation studies of SNRs show that young SNRs
have much lower polarisation fractions than old SNRs, see Fig.~\ref{fig:dickel} adopted from \citep{dickel90}.  Moreover, the magnetic field orientation in young SNRs tend to be
radially oriented, whereas for old SNRs the fields are tangential \citep{dickel76}. 
The tangential fields of old SNRs can be easily understood: the shock wave compresses only the magnetic field component
running perpendicular to the shock normal. The radial field orientation in young SNRs is less well understood; it may have something to do with  magnetic-field amplification in
the cosmic-ray precursors of SNRs \citep{zirakashvili08a}, or caused by a selection effect on where relativistic electrons pile up downstream of the shock \citep{west17}.
X-ray polarisation studies may shed some new light on this issue as it will allow for polarisation measurement close to the shock, and establish whether the
radial structure is already present there, or is caused by processes further away from the shock.

The low polarisation fraction of young SNRs in the radio, potentially makes the detection of X-ray polarisation challenging. On the other hand, since the X-ray synchrotron rims
are confined to regions close to  shocks fronts, X-ray polarisation studies offer the opportunity to study magnetic field orientations and turbulence close to
sites of active acceleration.
Related to this is that  the X-ray synchrotron emitting regions occupy less volume than the radio synchrotron emitting regions. So depolarisation due to different magnetic field orientations along
the line of sight is less likely in the X-rays than in the radio. So the X-ray polarisation fraction may be higher in X-ray than in radio. Finally, the maximum polarisation fraction depends on
the spectral index of the synchrotron emission. The maximum polarisation fraction depends on the spectral photon number index, $\Gamma$, as \citep[c.f.][]{ginzburg65}
\begin{equation}
\Pi_\mathrm{max}=\frac{\Gamma}{\Gamma + \frac{2}{3}}.
\end{equation}
So a steeper index will result in higher polarisation fractions.
Since the X-ray synchrotron emission is associated with photon energies near spectral cut-off, 
the spectral index is usually steeper than in the radio (note that $\Gamma = \alpha +1$, with $\alpha$ the radio spectral energy index).
In the radio young SNRs have typically $\Gamma\approx 1.6$, corresponding to $\Pi_\mathrm{max}\approx 70.5$\%, whereas in X-rays $\Gamma\approx 3$,
corresponding to $\Pi_\mathrm{max}\approx 82$\%.
In addition, the X-ray synchrotron emission may be selecting out more the high magnetic field peaks, where the higher magnetic fields push the cut-off energies into the X-ray band.
This effectively reduces the volume of the  X-ray synchrotron emitting plasma even further, enhancing the polarisation fraction.  All effects discussed so far
suggest that X-ray synchrotron polarisation fraction may be higher than radio polarisation. 
However, the proximity of X-ray synchrotron radiation regions to the shock front, may preferentially select regions with high magnetic-field turbulence.
So this could result in lower  polarisation fractions in X-rays as compared to the radio, for which the emission comes on average from regions further away from the shock front.
The magnitude of the magnetic-field fluctuations and their size distributions  has been described  quantitatively 
in \citep{bykov17}, who show that X-ray polarisation studies are an important means to study the magnetic field fluctuations that are so important for a proper understanding
of DSA in young SNRs.

In the next section we show what the detection possibilities are for polarisation in young SNRs, using the typical polarisation fractions based on the radio polarisation studies (Table~\ref{tab:radiopol}).

\begin{figure}
  \centering
  \includegraphics[trim=50 100 50 100,clip=true,height=0.3\textwidth]{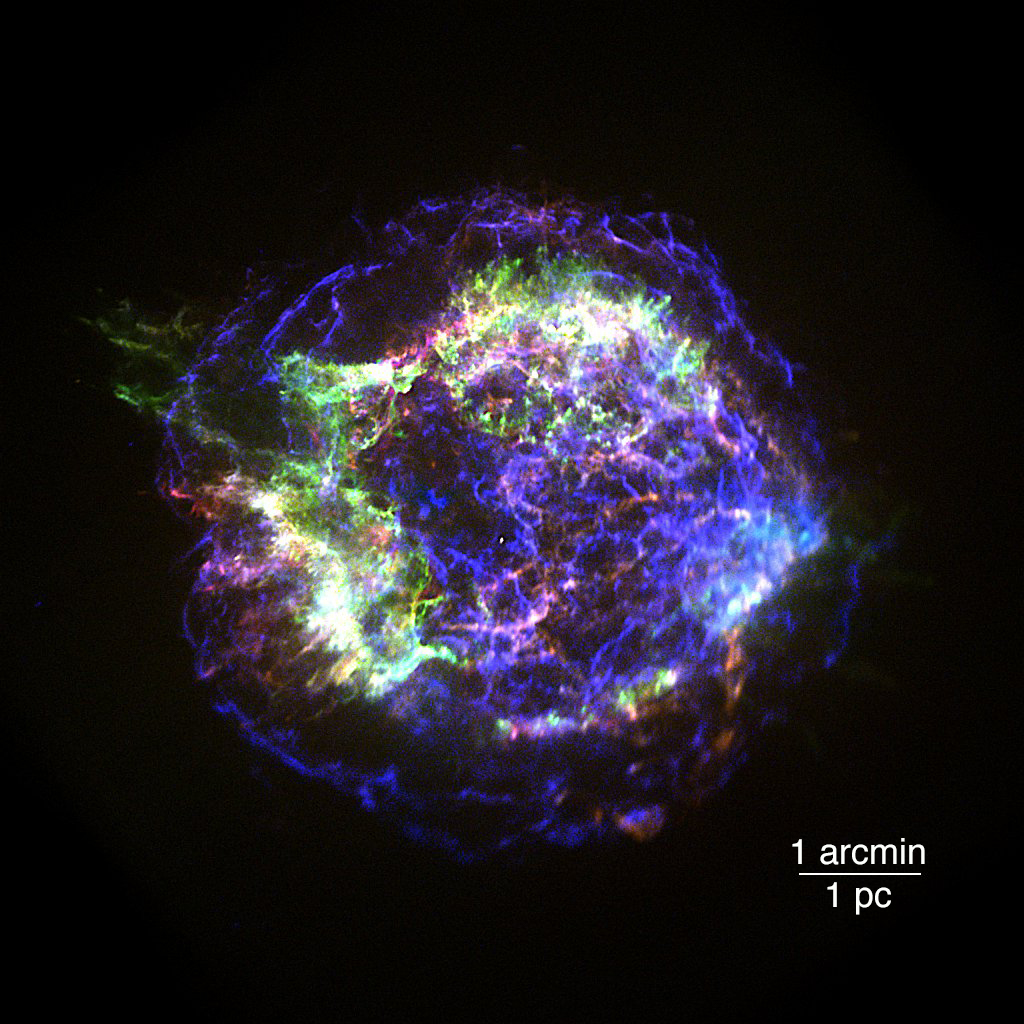}
    \includegraphics[trim=50 100 50 100,clip=true,height=0.3\textwidth]{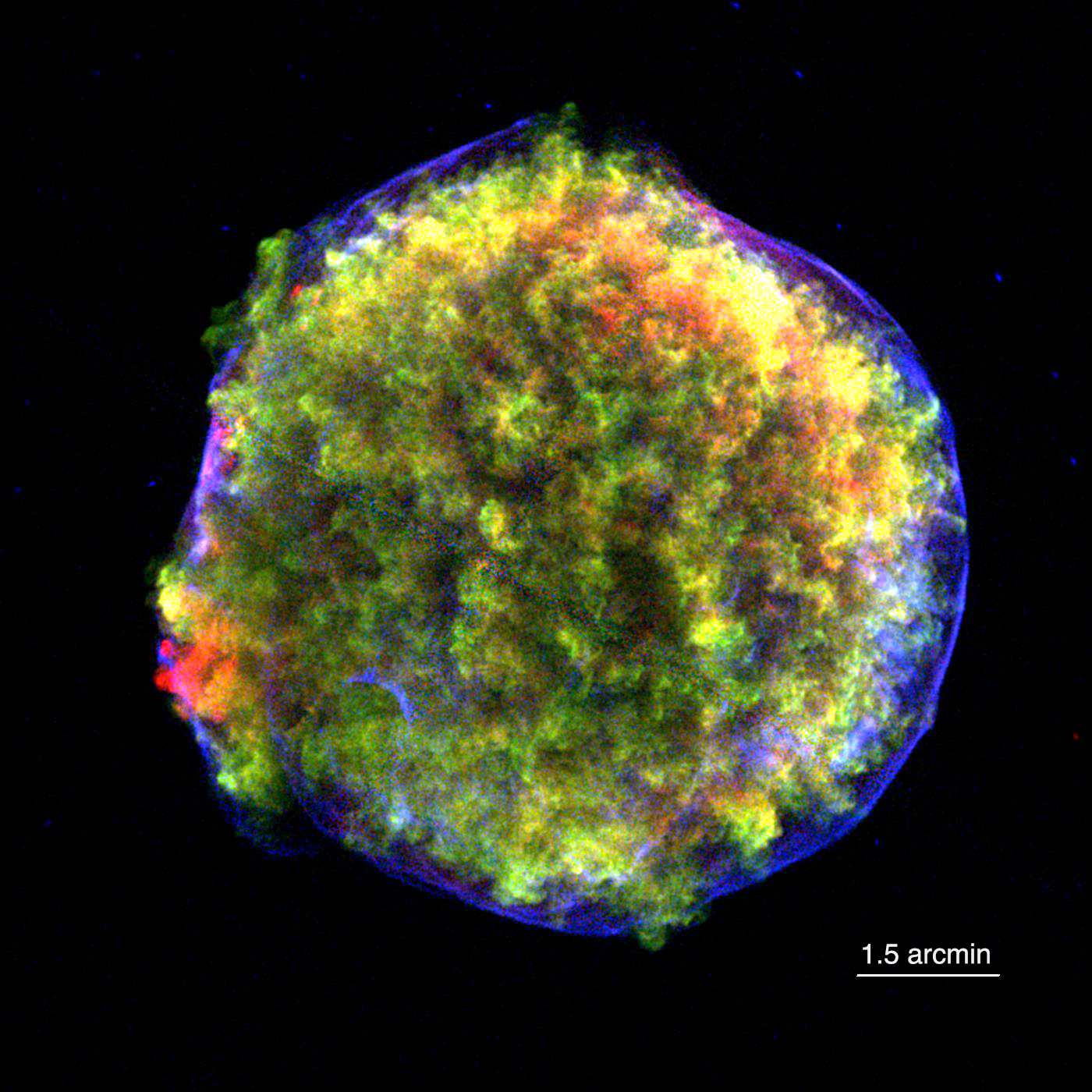}
   \includegraphics[trim=80 30 30 60,clip=true,height=0.3\textwidth]{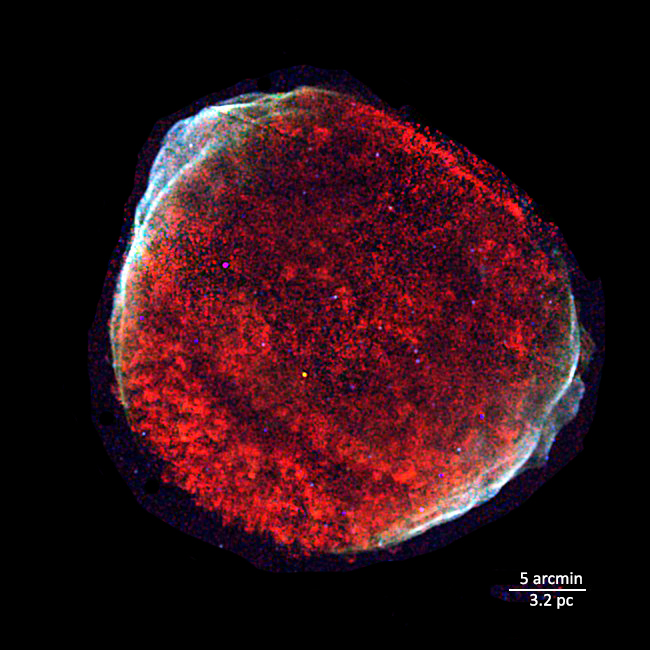}
   \caption{\label{fig:snrs}
  X-ray images of SNRs as observed by Chandra. In all cases the bluish colors represent  X-ray synchrotron dominated emission, which in principle should be polarised.
    Left: Cassiopeia A as observed in 2004.
    Red represents Mg XII and Fe L line
    emission, green Si XIII emission (around 1.85 keV), and blue is the
    4-6 keV continuum band.
    Middle: Tycho's SNR, blue is the 4-6 keV band,
        green is Si XIII (1.75-1.95 keV)and red is Fe L emission (0.8-0.95 keV).
    Right: SN\,1006, red is the 0.5-0.65 keV band, dominated by
    O VII oxygen emission, green is 0.85-2.0 keV, dominated by line emission from Ne IX, Mg XI and Si XIII (and lower ionisation states), and blue is 3-6.3 keV, dominated by synchrotron emission.
  }
\end{figure}

 \section{Simulations of supernova remnants}

The purpose of the set of simulations we will describe in this section is
twofold:  1) it should inform us on what what type of results
to expect from future
X-ray polarisation missions and what exposure times can realistically
expected to result in polarisation detection; and 2)
help us test and experiment with data analysis techniques.

For the input polarisation models we simply specified a location-dependent polarisation fraction and direction. For a discussion of the
polarisation effects
of underlying magnetic field turbulence on the X-ray
polarisation signal we refer to \citep{bykov17}. The two methods can be combined
as shown in Fig.~2-3 of the XIPE yellowbook.\footnote{\url{https://www.cosmos.esa.int/documents/1365222/1365271/SCI-2017-4+XIPE.pdf/1943687e-4d31-af62-ab79-1b0e1445ef14}.} 

We present here simulations of three SNRs: Cas A, Tycho (SN1572),
and SN\,1006. For the simulations we made educated guesses about what the
polarisation fraction may be in different locations of the remnants, based
on radio polarisation measurements (Table~\ref{tab:radiopol}). But we emphasise again that the X-ray synchrotron polarisation may be different, as regions close to shocks are sampled,
and the line-of-sights are shorter.
The simulations were done assuming the effective area and PSF of XIPE, which has characteristics similar to IXPE and eXTP. The input spectral parameters
are listed in Table~\ref{tab:input}.

 \begin{table}
  \caption{\label{tab:input}
  Spectral power-law parameters used for the simulations.}
  \centering
  \begin{tabular}{lllll}\hline\hline\noalign{\smallskip}
    Object & $\Gamma$ & norm & Band simulated & Reference \\
     & & & (ph\ s$^{-1}$\,keV$^{-1}$ @ 1 keV)& (keV) \\\noalign{\smallskip}\hline\noalign{\smallskip}
    Cas A   & 3.32 & 2.37 & 4- 6 &\citep{vink03a}\\
    SN 1572 & 3.2 & 0.77 & 4-6 & Based on Chandra data\\
    SN 1006 & 2.73 & 0.0405 & 2 - 8 & \citep{bamba08}\\
    \noalign{\smallskip}\hline
    \end{tabular}
\end{table}

 \subsection{Cas A}
 
Cas A is a $\sim 340$ year old SNR, has a diameter of 5.5\arcmin, and is one of the brightest
radio sources in the sky. Its X-ray synchrotron filaments are very narrow
$\sim 1-2$\arcsec \citep{vink03a} and, unique among SNRs, seems to
be associated both with the forward and the reverse shock
\citep{helder08,uchiyama08}.

The radio polarisation fraction of Cas A is unusually low, 5\%, and the magnetic
fields are radially oriented throughout the shell, but perhaps with some
deviations near
the forward shock \citep{gotthelf01a}.
A concern for X-ray polarisation measurements of Cas A is that the bright
shell and interior X-ray filaments smear out photons to the projected
locations of the narrow filaments at the forward shock.
In order to test whether this would wipe-out the polarisation signal at the
forward shock, we set-up the polarisation in a way that maximises the
depolarisation effects of the PSF: we assume that the magnetic field
at the outer filaments is perpendicular (polarisation vectors radial),
whereas for the interior region the polarisation vectors are perpendicular.
On the other hand we assume, perhaps somewhat optimistically, that the outer
filaments are characterised by a polarisation fraction of 15\%, whereas for
the interior filaments it is assumed to be 5\% (close to the radio polarisation fractions).

The input image was the Chandra 2004 image \citep{hwang04}
in the 4-6 keV continuum band (the
blue part in Fig.~\ref{fig:snrs}, left), and also the simulated output
is for that energy band. Given the narrow range in photon energy, the
effects of the different weighting schemes is small. 
The simulated exposure times were 2 Ms, necessary to obtain sufficient signal to noise
per pixel in the narrow energy band.

\begin{figure}
\centering
  \includegraphics[trim=0 0 0 0,clip=true,width=0.43\textwidth]{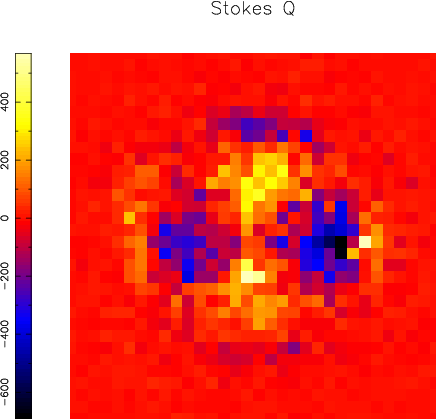}
    \includegraphics[trim=0 0 0 0,clip=true,width=0.43\textwidth]{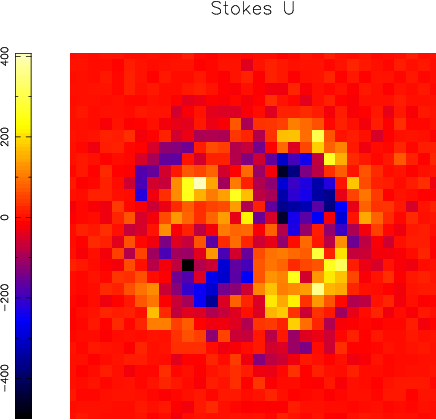}
  \includegraphics[trim=0 0 0 -40,clip=true,width=0.43\textwidth]{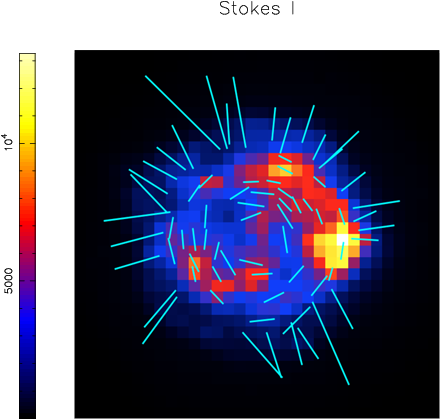}
    \includegraphics[trim=0 0 0 -40,clip=true,width=0.43\textwidth]{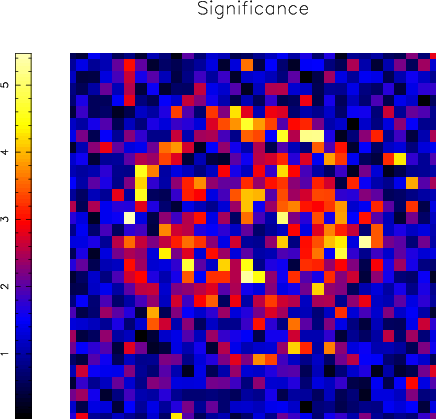}
  \includegraphics[trim=0 0 0 -40,clip=true,width=0.43\textwidth]{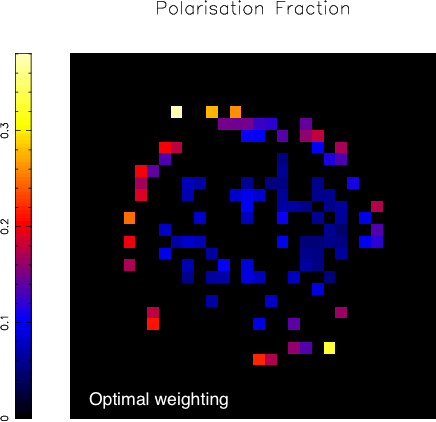}
    \includegraphics[trim=0 0 0 -40,clip=true,width=0.43\textwidth]{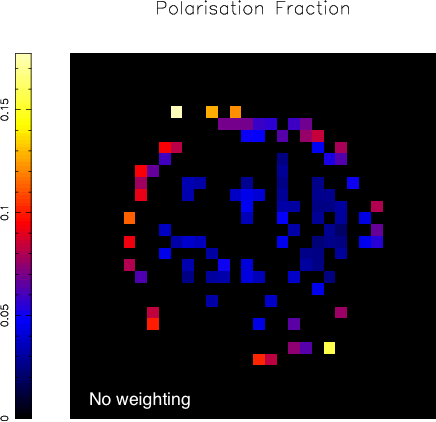}
  \caption{\label{fig:casa_mc}
  Monte Carlo simulations for Cassiopeia A based on 2 Ms observation with XIPE using photons in the 4-6 keV band. The pixel
  size is 16\arcsec.
  See the labels for the image identifications.  
  The Stokes I image shows the polarisation E-vector directions, with the  length of the line proportional to the
  polarisation fraction. For the polarisation fractions (lower two images) only pixels for which the polarisation signal to noise ratio is larger
  than 3 are indicated. Spurious "detections" outside the border of Cas A (see significance map) were ignored.
  All except the last image (lower right) are based on the weighted scheme for determining the stokes parameters. The last image gives
  the polarisation fraction without corrections. Visually it is similar to the image in the lower left, but its colour scaling is different.
  }
\end{figure}

\begin{figure}
\centering
  \includegraphics[trim=0 0 0 0,clip=true,width=0.43\textwidth]{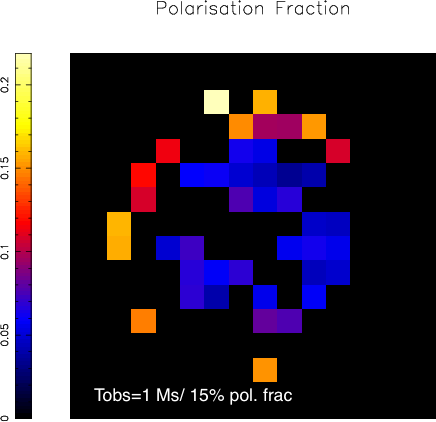}
  \includegraphics[trim=0 0 0 0,clip=true,width=0.43\textwidth]{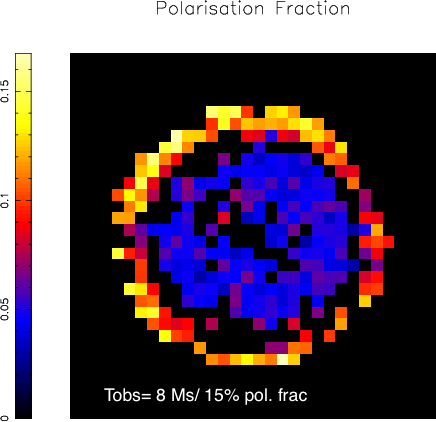}
   \includegraphics[trim=0 0 0 0,clip=true,width=0.43\textwidth]{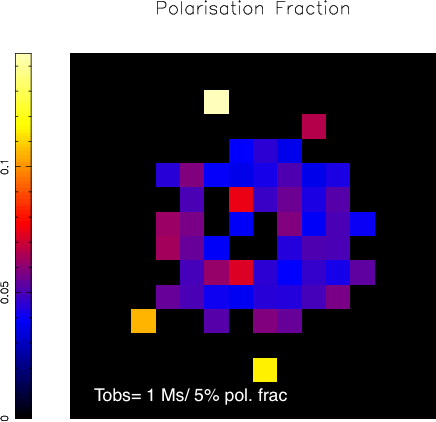}
      \includegraphics[trim=0 0 0 0,clip=true,width=0.43\textwidth]{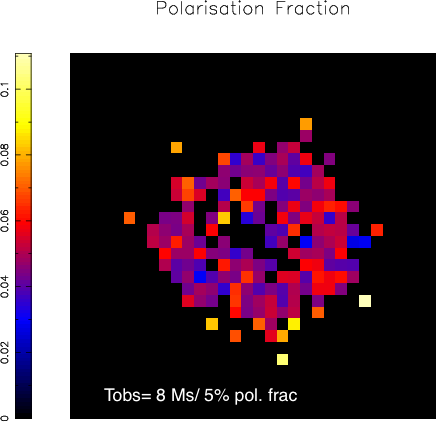}
\caption{
\label{fig:casa_mc_2}
Top:  Polarisation fractions for simulations of Cas A with 1 Ms (left, pixel size 32\arcsec) and 8 Ms (right) exposures.
Bottom: Simulated polarisation fractions, but now the outer filaments have input polarisation fractions of 5\%.
Both in the 2~Ms (left) and 8~Ms (right) the filaments are not detected as a coherent structure.
}
\end{figure}

As can be seen in Fig.~\ref{fig:casa_mc} the two different regions, outer filaments and inner filaments, imposed on
the input model, were nicely recovered, despite the perpendicular polarisation directions. The weighted scheme recovers
within the errors the input polarisation fractions. In Fig.~\ref{fig:casa_mc_2} a shallower and a deeper exposure are shown (top)
and simulations with a lower input polarisation fraction at the outer filaments (bottom).
The top left image shows that polarisation signals can still be detected with an exposure of 1~Ms, but only for a limited set of pixels and at
a courser binning level. Apparently the bright western inner part filaments hamper detection of polarisation of the outer shock in this direction.
The right image (8 Ms) shows that with enough statistics the measured polarisation fractions are close to the input values, except for the
western region again, where the PSF mixes signals from the inner and outer regions.
For lower polarisations levels of 5\% the PSF destroys the polarisation signal of the outer filaments. 

  \subsection{Tycho's SNR (SN 1572)}

Tycho's SNR has a diameter of 9\arcmin, which  still
fits within the field of view of the XIPE/IXPE and eXTP detectors.
Unlike for Cas A, the X-ray synchrotron emission  seems to be mostly
associated with the forward shock, forming a narrow filament marking the
shock front (the blueish/purplish structure Fig.~\ref{fig:snrs} centre).
Apart from the strikingly thin outer filament, there are a few peculiarly shaped
continuum features more toward the centre, which may be outer regions projected to the inside.
The best known, and most discussed structure is known as the "Tycho Stripes" on the western side, whose
regular structure have been suggested to represent some coherent magnetic field structure shaped by $10^{15}$~eV cosmic rays \citep{eriksen11}.
Another feature is an arc-like structure in the southeast.

\begin{figure}
\centering
\includegraphics[trim=0 0 0 0,clip=true,width=0.43\textwidth]{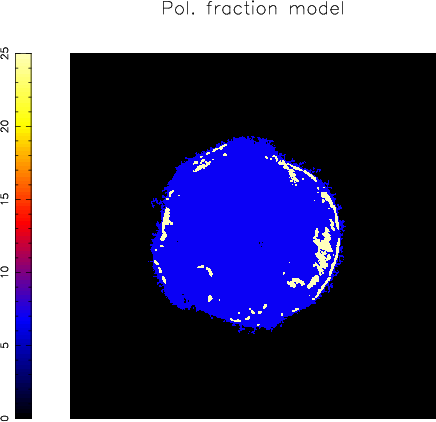}
\includegraphics[trim=0 0 0 0,clip=true,width=0.43\textwidth]{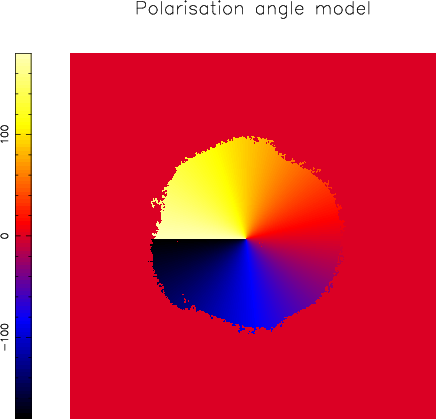}
\includegraphics[trim=0 0 0 0,clip=true,width=0.43\textwidth]{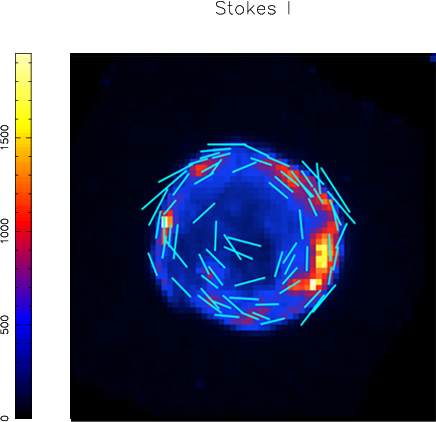}
\includegraphics[trim=0 0 0 0,clip=true,width=0.43\textwidth]{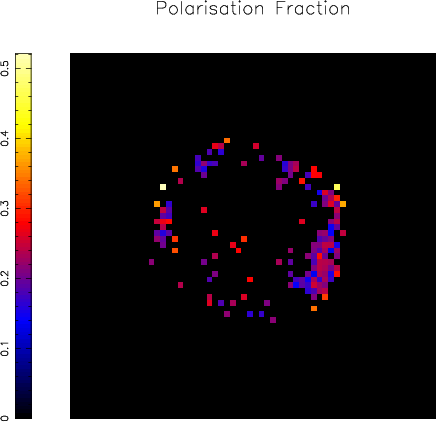}
\caption{\label{fig:tycho}
  Simulations using a toy input model for Tycho's SNR, based on the Chandra
  4-6 keV image  and input maps of polarisation fraction and angle,
  and assuming an observation time of 2~Ms.
  The output images are binned to 16\arcsec\ pixels.
  Top left: Input model for polarisation fraction (in \%).
  Top right: model for the polarisation angle (in degrees).
   Bottom, left: Simulated 4-6 keV count image (Stokes I). Bottom right: reconstructured polarisation fraction.
 }
\end{figure}

For this SNR we used the method of input maps ( Fig.~\ref{fig:tycho} top row) for the polarisation structure.
For our simulation of the 4-6 keV band,  we used the method of input maps ( Fig.~\ref{fig:tycho} top row) for the polarisation structure.
For the simulated observation time we used 2~Ms.
Based on the radio polarisation (Table~\ref{tab:radiopol}) most of the continuum was assumed to be 7\% polarised, but we assumed that 
the brightest X-ray continuum features have a polarisation fraction of 25\%. This includes a large part of the outer filament, as well as the stripes. The magnetic field was assumed
to be radially structured over the whole remnant (the polarisation vectors are then tangential).

As can be seen in Fig.~\ref{fig:tycho} for the brightest features the polarisation signal is recovered, but the weaker interior polarisation is not, except for regions
associated with the stripes and the arc. Even if we increase the exposure time to 8 Ms,
we cannot detect the interior, and a few high significant pixels are probably caused by scattering from high polarisation structures by the PSF.
It should be remembered, however, that the X-ray synchrotron emission from Tycho's SNR is highly limbbrightened and
the 4-6 keV continuum surface brightness of the interior is low.

 \subsection{SN 1006}
 
The X-ray synchrotron emission from SN~1006 comes from two X-ray limbs in the northeast and southwest \citep{koyama95}, respectively, 
each of which can be roughly covered with a single XIPE/IXPE pointing. The total diameter of SN\,1006 is 30\arcmin. 
Unlike for Cas~A and Tycho's SNR, the synchrotron emitting regions emit very little thermal radiation, and most emission above
1~keV is synchrotron emission, whereas from the limbs very little thermal radiation is emitted in the first place.
So X-ray polarisation studies do not have to be confined to the 4-6~keV continuum dominated band.
For that reason we set-up the Monte Carlo simulations of SN~1006 for the  2-- 8~keV band, using a 
power-law spectrum (Table~\ref{tab:input}) and an assumed polarisation fraction of 17\%, based
on radio observations (Table~\ref{tab:radiopol}). Since the effective area at 2~keV is larger than
at 4~keV, and a wider spectral coverage is used, the exposure time for each position was taken to be 1~Ms,
shorter than for Cas A and Tycho's SNR.

\begin{figure}
 \centering
 \includegraphics[trim=0 0 0 0,clip=true,width=0.32\textwidth]{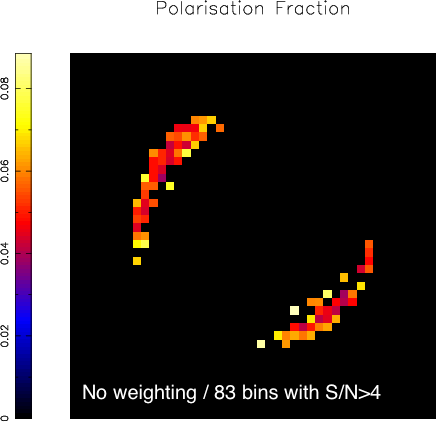}
 \includegraphics[trim=0 0 0 0,clip=true,width=0.32\textwidth]{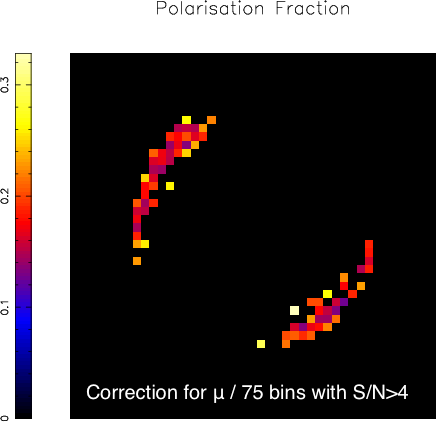}
 \includegraphics[trim=0 0 0 0,clip=true,width=0.32\textwidth]{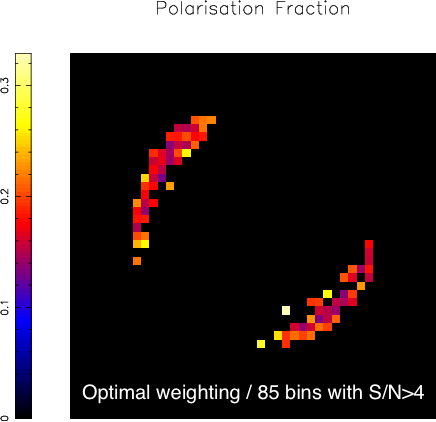}

 \includegraphics[trim=0 0 0 0,clip=true,width=0.45\textwidth]{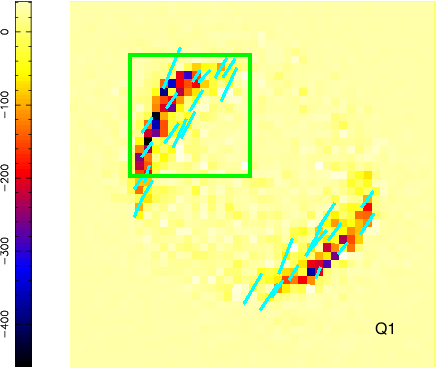}
 \includegraphics[trim=0 0 0 0,clip=true,width=0.45\textwidth]{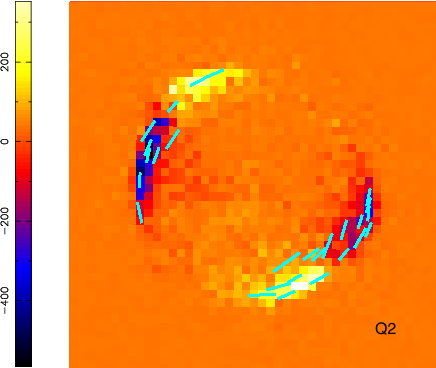}
  \includegraphics[trim=0 0 0 0,clip=true,width=0.45\textwidth]{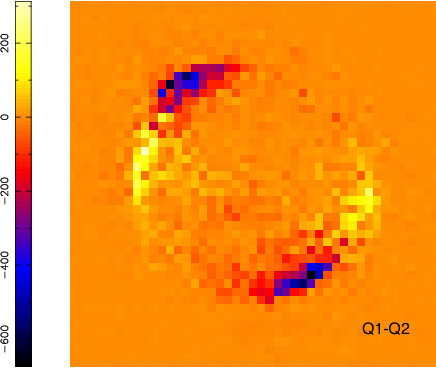}
    \includegraphics[trim=0 0 0 0,clip=true,width=0.45\textwidth]{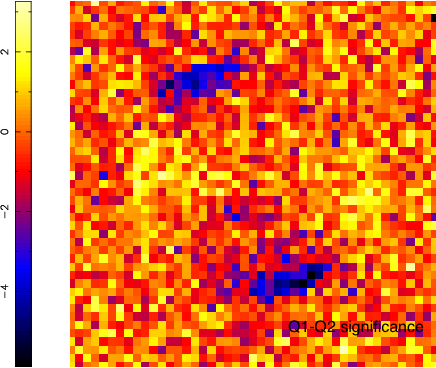}
  \caption{\label{fig:sn1006}
  Monte Carlo simulations of SN~1006 in 2-- 8 keV band based on 1 Ms exposure of each
  position with XIPE. The output images are binned to $64''$ pixels.
  Top panels: the polarisation fractions corresponding to the three weighting methods: 
 no weighting, correction for $\mu$ and an optimal weighting. A radial magnetic
 field is assumed for the two images.
  Middle panels: the simulated Stokes Q images using optimal weighting for parallel (left) and radial 
  magnetic field (right)
  configurations, respectively, overlaid with polarisation vectors (cyan bars) and the
  field of view of XIPE (green box). 
  Bottom panels: a difference map for the Q images in the middle panels and
  the significance map of the difference image on the left.
    }
\end{figure}

Given that a broader energy coverage can be used, SN\,1006 
is an ideal target to test how the weighting method (\S~\ref{sec:statistics}) affect the  measured results.
The upper panels of Figure~\ref{fig:sn1006} show the simulated polarisation fractions
of  SN~1006  assuming radial magnetic fields, using the three aforementioned weighting methods 
(left: no weighting; middle: corrected for $\mu$; right: an optimal weighting).
Only the data with S/N > 4 are shown in the images and a spurious data point was
removed.
Here the optimal weighting is found to provide more bins (85 pixels) with high polarisation significance (S/N>3--4)
than for the corrected method (75 significantly detected pixels).

The magnetic fields in SN~1006 are likely to be predominantly radial, but 
the synchrotron morphology has also been explained by invoking parallel magnetic field
lines oriented in the northeast-southwest direction 
\citep{rothenflug04,reynoso13}, reflecting the large scale magnetic fields at the location of SN\,1006.
These two magnetic field configurations (radial and parallel) can be significantly distinguished with a
1~Ms X-ray polarisation measurement, as shown in the middle and bottom panels of Figure~\ref{fig:sn1006}.
The simulated images of Stokes Q for parallel ($Q_1$) and radial ($Q_2$) B-configurations
reveal significantly different patterns, as also indicated by the the difference 
image $Q_1-Q_2$. The significance image of $Q_1-Q_2$ is obtained as
 $(Q_1-Q_2)/(\sqrt{{\rm Var}(Q_1)+{\rm Var}(Q_2))}$.

\section{Conclusion}

We have explored here both the science case for X-ray polarisation studies of young
SNRs, as well as exploring the practical issue of obtaining
X-ray polarisation maps with future instruments on board XIPE, IXPE and eXTP.

In particular, we advocated  to use estimators of the Stokes parameters based on
summations of the sine and cosine of the photon-electron ejection angles.
We expanded on the work of \citet{kislat15} by also including an energy dependent
correction factor, which corrects for instrumental effects, thereby providing estimates of the actual source 
polarisation fraction. The method is illustrated with simulations of X-ray polarisation maps
of young SNRs,
but it can also be used for X-ray polarisation timing and spectroscopy studies.

Our simulations show that long exposure times, $\sim1 - 2$~Ms, will be needed to
detect polarisation signals from young SNRs, but  for the important SNRs
Cas~A, Tycho's SNR and SN\,1006 the polarisation signals can be obtained for
polarisation fractions comparable to the radio detected polarisation fractions.
The long exposure times are  necessary to acquire sufficient polarisation statistics, 
which is more critical than for simple imaging. This is also due to the fact that
the GPD detectors have relatively small modulation factors and low quantum efficiencies 
\citep[less than 30\% above 2~keV,][]{bellazini06}.
So for X-ray polarisation measurements of astrophysical sources in general, longer
exposure times ($>100$~ks) are to be expected than for imaging spectroscopy observations
(typically a few 10~ks). 
The number
of sources to be observed during the span of the mission is, therefore, necessarily 
limited. Note that in the yellowbook for XIPE and its mock observation plan these
long observation times, including Ms observations of SNRs, were already taken into account. 

For SNRs
the additional complication is that one needs to rely on the 4-6~keV band to reliably isolate
the synchrotron continuum component, since below 4 keV line emission from
magnesium, silicon, sulphur, argon and calcium dominate the emission, and above $\sim 6.3$~keV
iron-K line emission becomes important. However, for isolated synchrotron filaments in certain
regions of Cas A and Tycho, one may be able to use the full X-ray band available
for polarisation studies (roughly 1.5-8 keV), or single out isolated
spectral regions with weak line contributions. As shown in this publication, for SN\,1006 one
can already use the 2-8~keV band, as the synchrotron dominated outer rims have only weak line emission.

The X-ray polarisation fractions adopted for our simulations are based on radio polarisation measurement,
but the X-ray polarisation fractions  may well be different, as the emitting
volumes are smaller and the spectral slopes are steeper
(both boosting the polarisation fraction), but also
comes from regions close to the shock front, where magnetic field turbulence
may be higher (which would lead to lower polarisation fractions).
The uncertainty on the X-ray polarisation fraction to be expected, as well as the impact
the detection of X-ray polarisation would have on our understanding of
magnetic field turbulence and particle acceleration efficiency in young SNRs, highlight the
need for X-ray polarisation missions.


\acknowledgments{P.Z. acknowledges the support from the NWO Veni Fellowship, grant no. 639.041.647
and NSFC grants 11503008 and 11590781.}

\authorcontributions{
JV has developed the xpolim X-ray polarisation simulation code, and has developed the mathematical equations. The simulations themselves were carried out by JV and PZ together.
}

\conflictsofinterest{``The founding sponsors had no role in the design of the study; in the collection, analyses, or interpretation of data; in the writing of the manuscript, and in the decision to publish the results''.} 

\abbreviations{The following abbreviations are used in this manuscript:\\

\noindent 
\begin{tabular}{@{}ll}
SNR(s)  & supernova remnant(s)\\
PWN(e) & pulsar wind nebula(e)\\
GPD & gas pixel detector\\
MDPI & Multidisciplinary Digital Publishing Institute\\
DOAJ & Directory of open access journals\\
TLA & Three letter acronym\\
LD & linear dichroism
\end{tabular}}



\end{document}